# Solution of the Deutsch-Josza Problem by NMR Ensemble Computing Without Sensitivity Scaling

F. M. Woodward  and  R. Brüschweiler[*]

Carlson School of Chemistry and Biochemistry, Clark University, Worcester, MA 01610

Correspondence to be addressed to:

Prof. Rafael Brüschweiler
Carlson Chair
Carlson School of Chemistry and Biochemistry
Clark University
Worcester, MA 01610-1477
Phone:       (508) 793-7220
Fax:         (508) 793-8861
E-Mail: bruschweiler@nmr.clarku.edu

## Abstract

Experimental NMR implementations of the Deutsch-Josza quantum algorithm based on pseudo-pure spin states exhibit an exponential sensitivity scaling with the number of qubits. By employing truly mixed spin states in spin Liouville space, where molecules with different nuclear spin configurations represent different input states, the Deutsch-Josza problem can be solved by a single function evaluation without a sensitivity loss concomitant with an increase of the number of bits.

Nearly all NMR quantum computations reported in the literature to date start out with a *pseudo-pure state* [1,2] represented by the density operator $\sigma = (1-\varepsilon)2^{-n}\mathbf{1} + \varepsilon|00\ldots0\rangle\langle00\ldots0|$, where $n$ is the number of spin 1/2 nuclei (= quantum bits or qubits) of a single molecule of the ensemble, $|00\ldots0\rangle$ is usually the spin-wave function of the ground state of one molecule, and $\mathbf{1}$ is the unity operator (for a recent review see e.g. Ref. [3]). Despite its formal similarity to single pure state quantum computers, the actual "quantumness" of such computations can be affected [4-6]. The prefactor $\varepsilon$, which determines the sensitivity, decreases exponentially with the number of qubits, which renders this approach practical only for a restricted number of qubits.

Modern liquid-state NMR, with its highly developed pulse methods for generating a large variety of unitary evolutions [7], offers alternative schemes to perform certain computational tasks efficiently. Recently, a formalism for NMR quantum computations was introduced that is based on direct products of spin-polarization operators to represent logic states [8]. Superpositions of these states are generally truly mixed, i.e. they cannot be represented in spin Hilbert space but rather in Liouville space, which spans all conceivable spin density operators $\sigma$ characterizing the spin ensemble. This kind of computational strategy does not fit a narrow definition of "quantum computation" [6], although the involved scalar spin-spin coupling Hamiltonian has a genuinely quantum-mechanical nature. Unlike computations using pseudo-pure states, this scheme does not necessarily suffer from an exponential loss of sensitivity nor is it bound to the speedup limits of pure-state quantum computers [9]. It is shown here that the Deutsch-Josza problem can be solved with a sensitivity that does not scale with the number of qubits.

The $2^n$ eigenstates $|\psi\rangle$ of the Zeeman Hamiltonian, created by a strong external magnetic field, have the form $|\psi\rangle = |001\ldots01\rangle = |\alpha\alpha\beta\ldots\alpha\beta\rangle$ ($\alpha$ denotes spin "up" and $\beta$ spin "down"). These states can be mapped on states in spin Liouville space [8]

$$|\psi_{in}\rangle = |\alpha\alpha\beta\ldots\alpha\beta\rangle \quad \Rightarrow \quad \sigma_{in} = |\psi_{in}\rangle\langle\psi_{in}| = I_1^\alpha I_2^\alpha I_3^\beta \ldots I_{n-1}^\alpha I_n^\beta \quad (1)$$

where $I_1^\alpha I_2^\alpha I_3^\beta \ldots I_{n-1}^\alpha I_n^\beta$ is the direct product of *polarization operators* $I_k^\alpha, I_k^\beta$ defined as [7]

$$I_k^\alpha = |\alpha_k\rangle\langle\alpha_k| = \frac{1}{2}(I_k + 2I_{kz}) = \begin{pmatrix} 1 & 0 \\ 0 & 0 \end{pmatrix}, \quad I_k^\beta = |\beta_k\rangle\langle\beta_k| = \frac{1}{2}(I_k - 2I_{kz}) = \begin{pmatrix} 0 & 0 \\ 0 & 1 \end{pmatrix}. \quad (2)$$

$2I_{kz}$ is the Pauli matrix $\sigma_z$ and $I_k$ the unity operator of the subspace of spin $I_k$. It is useful to note that $I_k^\alpha + I_k^\beta = I_k$ and $I_k^\alpha - I_k^\beta = 2I_{kz}$. $\sigma_{in}$ has the dimension $2^n$ and its matrix elements in the Zeeman basis are all zero except for one diagonal element belonging to state $|\alpha\alpha\beta\ldots\alpha\beta\rangle$ which is 1. $I_k^\alpha$ is assigned to the logic state 0 while $I_k^\beta$ is assigned to the state 1.

A hallmark of computations in spin Liouville space is the ability to use as input state a linear combination (superposition) of different "classical" input states, represented by different molecular subensembles constituted of molecules residing in the same pure state, with the output corresponding to the superposition of the individual output states [8].

Due to its conceptual simplicity, the Deutsch-Josza problem [10] was one of the first quantum computational problems that were implemented using NMR spectroscopy [11-18]. Its task is to determine whether the Boolean function $f: \{0, 1\}^n \to \{0, 1\}$ is *constant*, always returning 0 or always 1 irrespective of the input, or *balanced* returning 0 for half of the inputs and 1 for the other half.

Functions of this kind are often referred to as "black box" or "oracle" and the details of their physical realization are not important here. It is sufficient to require that $f$ is linear

$$f(c_1\sigma_1 + c_2\sigma_2) = c_1 f(\sigma_1) + c_2 f(\sigma_2) \tag{3}$$

and to note that a reversible implementation of $f$ is possible by using (at least) one extra bit to represent the computation as a permutation of the states of the input bits [19]. The extra bit, represented by spin $I_0$, is at the beginning of the computation in a well-defined state such as the $I_0^\alpha$ state. The output is displayed on spin $I_r$: if $f = 0$ then its state is $I_r^\alpha$, if $f = 1$ its state is $I_r^\beta$.

The Deutsch-Josza quantum algorithm gives the correct answer with only one evaluation of $f$ while a classical algorithm requires $2^{n-1} + 1$ evaluations in the worst case [10]. NMR implementations of the Deutsch-Josza algorithm using pseudo-pure states are accompanied by the exponential sensitivity loss mentioned above. In Ref. [15] it was noted that the computation can be performed with the equilibrium density operator as input instead of a pseudo-pure state, the unfavorable scaling of the sensitivity, however, remained essentially unchanged.

We now use the linearity of $f$ mentioned above and evaluate $f$ on a superposition of input density operators of the kind of Eq. (1). In particular, if we prepare a superposition of all possible input states with a uniform weighting, we obtain the unity density operator times $I_0^\alpha$: $\sigma = I_0^\alpha \sum I_1^{\alpha/\beta} I_2^{\alpha/\beta} \ldots I_n^{\alpha/\beta} = I_0^\alpha \mathbf{1}$. This fact illustrates a fundamental difference between Liouville space and Hilbert space quantum computing: in spin Liouville space the state of "maximal superposition" has maximal entropy, while in Hilbert space entropy is minimal and constant. Experimental preparation of the state $\sigma = I_0^\alpha \mathbf{1}$ is straightforward: at thermal equilibrium, a hard $90^\circ$ pulse is applied to all input spins (but not to spin $I_0$) which is followed by a field-gradient "crusher" pulse (PFG). $I_0^\alpha \mathbf{1}$ remains invariant under any balanced function $f$, since $UI_0^\alpha \mathbf{1} U^\dagger = I_0^\alpha \mathbf{1}$ for all unitary transformations, including the permutation corresponding to $f$, applied in the subspace spanned by spins $I_1, \ldots, I_n$ and leaving $I_0$ unaffected. Since NMR observables are traceless operators, no detectable signal will arise from any of the spins $I_1, \ldots, I_n$ including the detection spin $I_r$. The result of the constant function, on the other hand, always yields a positive or always a negative signal on the detection spin depending whether $f$ is constant 0 or 1. (A conceivable implementation uses a copy operation (FANOUT) of $I_0^\alpha$ on the detection spin $I_r$ followed by an inversion pulse on $I_r$ if the constant output is $I_r^\beta$). Since the sensitivity does not depend on the number $n$ of input bits, the Deutsch-Josza problem is solvable in Liouville space with a single scan NMR experiment with a sensitivity that does not decrease with an increase of the number of qubits (spins) per molecule. This feature is quite different from the pseudo-pure state implementation of the Deutsch-Josza algorithm and its refined version [11-15]. With respect to both efficiency and scaling, the Liouville-space implementation is equivalent to the Deutsch-Josza algorithm performed on a pure-state quantum computer. Consequently, ensemble quantum computing is not necessarily accompanied by an exponential scaling of the sensitivity.

## Acknowledgment

Valuable discussion with Prof. Frederick Green is acknowledged. J. M. Myers, A. F. Fahmy, S. J. Glaser, and R. Marx developed a formal theory of NMR quantum computing and reached the same conclusion described here.